\documentclass[%
 twocolumn,
 amsmath,amssymb,
 aps,
]{revtex4}

\usepackage{graphicx}
\usepackage{dcolumn}
\usepackage{bm}
\usepackage{verbatim}
\usepackage[caption=false]{subfig}
\usepackage{floatrow}
\usepackage{silence}
\usepackage{color}

\newcommand{\bra}[1]{\ensuremath{\left\langle#1\right|}}
\newcommand{\ket}[1]{\ensuremath{\left|#1\right\rangle}}
\newcommand{\EEE}{\mathcal{E}}
\newcommand{\MMM}{\mathbf{M}}
\newcommand{\DDD}{\mathcal{D}}
\newcommand{\HHH}{\mathcal{H}}

\newcommand{\aaa}{\textbf{a}}
\newcommand{\bbb}{\textbf{b}}
\newcommand{\sss}{\boldsymbol{\sigma}}
\newcommand{\rrr}{\boldsymbol{\rho}}

\begin{document}


\title{Dissipation-induced continuous quantum error correction for superconducting circuits}

\author{Joachim Cohen$^{1,2}$}
 \email{joachim.cohen@inria.fr}
\author{Mazyar Mirrahimi$^{1,3}$}%
 \email{mazyar.mirrahimi@inria.fr}
\affiliation{%
 $^{1}$INRIA Paris-Rocquencourt, Domaine de Voluceau, Bo\^ite Postale 105, 78153 Le Chesnay Cedex, France\\
 $^{2}$Laboratoire Pierre Aigrain, Ecole Normale Supérieure, CNRS (UMR 8551), Université P. et M. Curie, Université D. Diderot 24, rue Lhomond, 75231 Paris Cedex 05, France\\
 $^{3}$Department of Applied Physics, Yale University, New Haven, CT 06520, USA\\
}%


\begin{abstract}
Quantum error correction (QEC) is a crucial step towards long coherence times required for efficient quantum information processing (QIP). One major challenge in this direction concerns the fast real-time analysis of error syndrome measurements and the associated feedback control. Recent proposals on autonomous QEC (AQEC) have opened new perspectives to overcome this difficulty. Here, we design an AQEC scheme based on quantum reservoir engineering adapted to superconducting qubits. We focus on a three-qubit bit-flip code, where three transmon qubits are dispersively coupled to a few low-Q resonator modes. By applying only continuous-wave drives of fixed but well-chosen frequencies and amplitudes, we  engineer an effective interaction Hamiltonian to evacuate the entropy created by eventual bit-flip errors. We provide a full analytical and numerical study of the protocol, while introducing the main limitations on the achievable error correction rates.
\end{abstract}

\maketitle


\section{\label{sec:level1}Introduction}

An essential requirement for the development of QIP is the active QEC~\cite{nielsen00}. By designing an encoded logical qubit, possibly using many physical qubits, one protects the quantum information against major decoherence channels and hence ensures a significantly longer coherence time than a physical qubit~\cite{Shor-QEC,Steane-PRL_1996}. A standard measurement-based feedback procedure to perform active QEC consists of probing some observables~\cite{nielsen00,Gottesmanthesis}, e.g. multi-qubit parities, in a non-destructive and repeated manner. Analyzing in real-time the measurement output reveals the occurrence of possible errors which could then be corrected by applying an appropriate unitary action in feedback. Recent advances in quantum-limited amplification~\cite{Lehnert-NatPhys-2008,Bergeal-Nat-2010,Hatridge-PRB-2011,Roch-et-al-12} have opened doors to high-fidelity non-demolition measurement of superconducting qubits and have already led to successful experiments on closed-loop control of such systems~\cite{vijay-et-al-2012,riste-PRL-2012,campagne-PRX-2013,deLange-PRL-2014}. However, the relevant time-scales for these systems impose important limitations on the complexity of real-time analysis that one can perform on the measurement output. In particular, the finite bandwidth of the amplification procedure, together with the time-consuming data acquisition and post-treatment of the output signal, lead to an important latency in the feedback procedure. 

Alternatively, the reservoir (dissipation) engineering~\cite{poyatos-et-al-96} and the closely related coherent feedback~\cite{Lloyd-PRA-2000} circumvent the necessity of a real-time data acquisition, signal processing and feedback calculation.  Coupling the quantum system to be stabilized to a strongly dissipative ancillary quantum system allows one to evacuate the entropy of the main system through the dissipation of the ancillary one. By building the feedback loop into the Hamiltonian, this type of autonomous feedback obviates the need for a complicated external control loop to correct errors. On the experimental side, such autonomous feedback techniques have been used for qubit reset~\cite{Geerlings-PRL-2013}, single-qubit state stabilization~\cite{PhysRevLett.109.183602}, and the creation~\cite{Barreiro-Nature-2011} and stabilization~\cite{Krauter-PRL-2011,Shankar-Nature-2013,Lin-Nature-2013} of states of multipartite quantum systems.

AQEC with multi-qubit codes has been theoretically investigated in a few recent proposals adapted to quantum photonics systems~\cite{PhysRevLett.105.040502,2011NJPh...13e5022K}. The approach of~\cite{PhysRevLett.105.040502,2011NJPh...13e5022K} consists in applying  an embedded optical feedback loop for the QEC where each qubit is coupled to a different optical resonator, and the directional coupling between these subsystems is ensured through waveguide connections. Here instead, we exploit the strong couplings and nonlinearities provided by quantum superconducting circuits to introduce important hardware shortcuts and to propose a protocol adapted to state of the art experiments in this context. More precisely, by considering three transmon qubits~\cite{Koch-et-al-07} coupled, in the strong dispersive regime~\cite{schuster-nature07}, to three (or one in a simplified version) low-Q modes of a single 3D cavity, we propose an AQEC protocol: by applying some appropriate Continuous-Wave (CW) microwave drives, we produce an effective Hamiltonian that evacuates the entropy resulting from bit-flip errors. 

The scheme being only based on the application of CW drives of fixed frequencies, amplitudes and phases (no time-dependence for these parameters), we ensure a strong robustness with respect to small variations of these parameters and require only basic experimental calibrations. Also, compared to the protocols in~\cite{PhysRevLett.105.040502,2011NJPh...13e5022K}, we avoid any requirement of directional couplings which greatly simplifies the experimental implementation of such a protocol with superconducting circuits. Indeed, ensuring any directionality  in the transmission of  quantum information, while avoiding corruption with extra noise, necessitates the development of  new quantum-limited devices based on Josephson elements and represents, by itself, a significant experimental objective. Moreover, in a similar manner to the recent work~\cite{PhysRevA.88.023849,Shankar-Nature-2013,DiVicenzo1,PhysRevA.89.032314,Roch-PRL-2014}, our protocol is based on minimal symmetry requirements: we only need a certain linear combination of the dispersive shift strengths to be small. Such a symmetry can be rather easily achieved by tuning the qubits frequencies (using for example double-junction qubits and applying external magnetic flux).  Finally, by avoiding resonant interactions between the qubits and the low-Q resonators, the qubits remain protected against the Purcell effect. 

In Section~\protect\ref{sec:framework}, we provide the framework of the AQEC scheme. After a brief overview of the idea behind the reservoir engineering for QEC, we introduce the considered physical system, together with the required coupling regimes.  The Section~\protect\ref{sec:level3} provides the AQEC protocol. In Subsection~\protect\ref{sec:level3A}, we present the idea on a simpler case where only one of the three qubits can undergo a bit-flip error. In Subsection~\protect\ref{sec:level3B}, we generalize the idea to the complete case where the three qubits suffer independently from bit-flip errors.  In Subsection~\protect\ref{ssec:summary}, we summarize the ideas and perform numerical simulations that illustrate the performance of the scheme with realistic experimental parameters. In Section~\protect\ref{sec:level4}, we expose the limitations of the proposed protocol through the analysis of major decoherence channels created by various possible imperfections. Finally, the Section~\protect\ref{sec:level5} is devoted to a simplified version of the protocol where we only require the coupling of the three qubits to a single low-Q resonator: this could be considered as the minimal experimental setup required for realizing a bit-flip code.

\section{Framework of autonomous QEC}\label{sec:framework}

\subsection{Reservoir engineering for QEC}\label{ssec:res}

The 3-qubit bit-flip code consists of encoding the logical states $\ket{0}$ and $\ket{1}$ using the states $\ket{000}$ and $\ket{111}$ of three  physical qubits. Starting from a superposition in the code space $\EEE_0=\text{span}\{\ket{000},\ket{111}\}$, a single bit-flip error maps the states to one of the \textit{error subspaces} $\EEE_1=\text{span}\{\ket{100},\ket{011}\}$, $\EEE_2=\text{span}\{\ket{010},\ket{101}\}$ or $\EEE_3=\text{span}\{\ket{001},\ket{110}\}$.  We can associate to these error processes, the Kraus operators $\MMM_0=\sqrt{1-p}\textbf{I}$, $\MMM_1= \sqrt{\frac{p}{3}}\sss_x^1$, $\MMM_2= \sqrt{\frac{p}{3}}\sss_x^2$ and $\MMM_3= \sqrt{\frac{p}{3}}\sss_x^3$, where $p\ll 1$ is the bit-flip probability for a single physical qubit, $\textbf{I}$ is the identity on the qubits Hilbert space, and $\sss_x^k$ is the Pauli matrix along the $X$ axis of the $k$'th qubit. 

In conventional QEC, a measurement of the two-qubit parities  would reveal the error subspace the system lives in without leaking out any further information on the superposition between the logical states. The quantum state could then be restored by applying an appropriate quantum gate.  Alternatively, in a reservoir engineering scheme, we use the coupling to an ancillary quantum system to mediate the evacuation of the information entropy. More precisely, we design a joint unitary operation $U_{SA}$ between the system (Hilbert space $\HHH_S$) and the ancilla (Hilbert space $\HHH_A$) satisfying
\begin{align*}
U_{SA}\left((\MMM_j\ket{000}_S)\otimes \ket{0}_A\right)&=\ket{000}_S\otimes \ket{j}_A,\\
U_{SA}\left((\MMM_j\ket{111}_S)\otimes \ket{0}_A\right) &=\ket{111}_S\otimes \ket{j}_A,~~ j=0,1,2,3.
\end{align*}
While the system is already projected back onto the code space $\EEE_0$, a rapid decay of the ancilla resets its state to $\ket{0}_A$, preparing it for the next run of QEC. Through our scheme (see Section~\protect\ref{sec:level3}), using driven damped harmonic oscillators as ancillary system, we perform these steps of unitary operation and ancilla reset in a continuous and simultaneous manner. 

\subsection{Physical system}\label{ssec:system}

\begin{figure}
	\centering
		\includegraphics[scale=0.27]{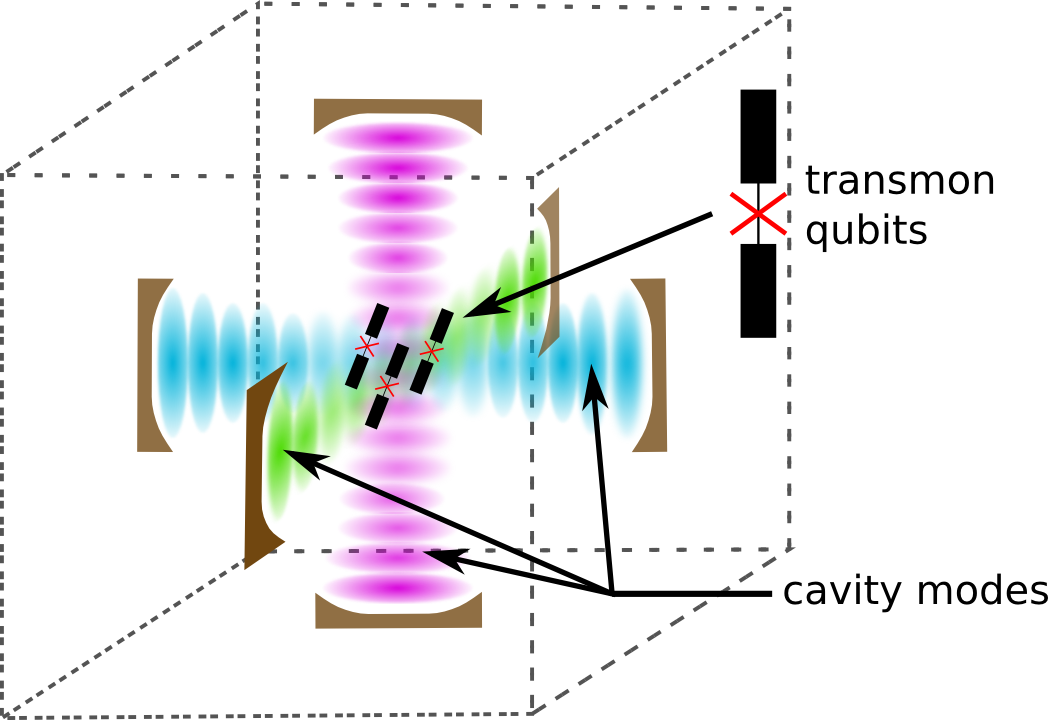}
	\caption[fancyfig]{A possible physical realization where three transmon qubits are strongly coupled to three low-Q spatial modes of a 3D superconducting cavity. The coupling of the qubits to the modes are designed such that~\eqref{eq:strongdisp} and~\eqref{eq:deg} are satisfied. External microwave drives may be applied to the cavity, but the output ports of the cavity are not monitored.}
	\label{fig:fancyfig}
\end{figure}

We consider three transmon qubits~\cite{Koch-et-al-07} coupled to three low-Q resonators. In Figure~\ref{fig:fancyfig}, we present a design where the three qubits are coupled to three spatial modes of a 3D superconducting cavity. While the qubits are used to encode the quantum information, the resonators together play the role of the ancilla. Following the strategy of the previous subsection, we will map the error subspaces $\EEE_{0,1,2,3}$ to the three ancilla states $\ket{000}_A$, $\ket{100}_A$, $\ket{010}_A$ and $\ket{001}_A$, where $\ket{0}_A$ and $\ket{1}_A$ are respectively the vacuum state and the single-photon Fock state of each resonator. 

The total Hamiltonian of the driven system can be written in the following form~\cite{PhysRevLett.108.240502},
\begin{align}\label{eq:Ham1}
 \bold{H}(t)&= \sum_{k=1}^3 \tilde\omega_{a_k} \aaa_k^\dag \aaa_k+\sum_{k=1}^3 \tilde\omega_{b_k} \bbb_k^\dag \bbb_k \notag\\
&- \sum_{k=1}^3 E_{J}^k \left(\cos\left(\frac{\bold{\Phi}_k}{\phi_0}\right)+\frac{1}{2}\frac{\bold{\Phi}_k^2}{\phi_0^2}\right)\notag\\
&+ \sum_{k=1}^3 \epsilon^a_k(t)(\aaa_k+\aaa_k^\dag)+\sum_{k=1}^3 \epsilon^b_k(t)(\bbb_k+\bbb_k^\dag), 
\end{align}
where 
$$
\bold{\Phi}_k=\sum_{k'=1}^3 \phi^a_{k,k'}(\aaa_{k'}+\aaa_{k'}^\dag)+\sum_{k'=1}^3 \phi^b_{k,k'}(\bbb_{k'}+\bbb_{k'}^\dag).
$$
Here we note $\aaa_k$ (resp. $\aaa_k^{\dagger}$) and $\bbb_k$ (resp. $\bbb_k^{\dagger}$) the 
annihilation (resp. creation) operator of resonator $k$ and qubit $k$, $\tilde\omega_{a_k}$ and $\tilde\omega_{b_k}$ the dressed frequencies
of resonator $k$ and qubit $k$ respectively, $E_{J}^k$ the Josephson energie of qubit $k$, $\phi_0 = \hbar/2e$ the superconducting quantum flux. Some external drives, denoted by $\epsilon_k^{a,b}(t)$, may also be applied to the resonators and the qubits.
Noting that $\phi^a_{k,k'}\ll\phi^b_{j,j}$, the dressed modes $\aaa$ share a much smaller part of the non-linearity than the  dressed modes $\bbb$. This is why we refer to the $b$ modes as the qubit modes and the $a$ modes as the cavity modes. 

In the transmon regime $\mid \frac {{\bold{\Phi}^k}}{\phi_0} \mid \ll 1$ and therefore we can neglect higher than fourth order terms in the cosines. In the absence of external drives and restricting ourselves to the first two levels of the qubit modes $\bbb_{1,2,3}$, the effective Hamiltonian, in the dispersive coupling regime (where the resonance frequencies are well separated), becomes~\cite{PhysRevLett.108.240502}
\begin{align}
 \label{eq:bareH}
 \widetilde H(t) & = \sum_{k=1}^3 \omega_{a_k} \aaa_k^\dag \aaa_k+\sum_{k=1}^3 \frac{ \omega_{b_k}}{2}\sss_z^k \notag\\
 & - \sum\limits_{k=1,2,3} {{\aaa_k}^{\dagger}\aaa_k}(\frac{\chi^{ab}_{k1}}{2}{\sss^1_z}+\frac{\chi^{ab}_{k2}}{2}{\sss^2_z}+\frac{\chi^{ab}_{k3}}{2} 
 {\sss^3_z}\notag)\\
 & - \sum_{k} \chi_{kk}^{aa}{{\aaa_k}^{\dagger 2}\aaa_k^2}- \sum_{j\neq k} \chi_{jk}^{aa }{{\aaa_j}^{\dagger}\aaa_j}{{\aaa_k}^{\dagger}\aaa_k}\notag\\
 &-  \sum_{j\neq k} \chi_{jk}^{bb} {\sss^j_z}{\sss^k_z}. 
\end{align}
In the above expression, we make use of the renormalized frequencies $\omega_{a_k} $ and  $\omega_{b_k}$ for the resonators modes and for the qubits modes.
 While the dispersive coupling strengths $\chi_{kj}^{ab}$ ($j,k=1,2,3$) are the key parameters in our QEC protocol, the other self-Kerr and cross-Kerr terms $\chi_{kk}^a$, $\chi_{jk}^a$ and $\chi_{jk}^b$ are small compared to the these dispersive couplings, as they represent higher order effects. However, as it will be seen later, our QEC protocol is fully insensitive to the contribution of these terms. 
 
Similarly to~\cite{PhysRevA.88.023849}, we consider the strong dispersive coupling regime, where the dispersive shifts $\chi_{j}^k$ are much larger than the qubit and the cavity linewidths:
\begin{equation}\label{eq:strongdisp}
|\chi^{ab}_{kj}| \gg \kappa_{k'},\gamma_{j'},\quad j,k,j',k'=1,2,3,
\end{equation}
where $\gamma_j$ and $\kappa_k$ represent, respectively, the linewidths of qubit $j$ and resonator $k$. An additional symmetry assumption 
$$
\sum_j \chi_{kj}^{ab}=0 \quad k=1,2,3,
$$
is required to ensure that the QEC protocol does not reveal any information other than the error syndromes. In practice, a finite sum of the dispersive couplings would lead to an extra dephasing between the code states $\ket{000}$ and $\ket{111}$ which could be neglected in the limit 
\begin{equation}\label{eq:deg}
\mid \sum_j \chi_{kj}^{ab}\mid \ll \kappa_k.
\end{equation} 
The assumptions~\eqref{eq:strongdisp} and~\eqref{eq:deg} imply that at least one of the coupling strengths $\chi_{kj}^{ab}$ for each resonator $k$ is negative. While this is considered to be a rather hard task for a transmon qubit to change the sign of its dispersive couplings, the above requirement could be relaxed by encoding the quantum information in a different subspace than $\text{span}\{\ket{000},\ket{111}\}$. Indeed, using the subspace $\text{span}\{\ket{100},\ket{011}\}$ instead, we rather need to satisfy $|\chi_{k1}^{ab}-\chi_{k2}^{ab}-\chi_{k3}^{ab}|$ to be small, which could be satisfied even for positive-valued $\chi_{kj}^{ab}$'s.

\section{\label{sec:level3}Error Correction Scheme}

In this Section, we describe in details the error correction scheme using three qubits coupled to three cavities. In a first subsection, we focus on a simple case where only one of the three qubits can undergo a bit-flip and therefore the correction takes place only on this qubit. Next, we will extend the  protocol to the case where the three qubits suffer from bit-flips. 

\subsection{\label{sec:level3A}Correction on one qubit} 

Through the rest of this paper, we consider the system in the rotating frame given by the Hamiltonian 
$$
\bold{H}_0=\sum_{k=1}^3 \omega_{a_k} \aaa_k^\dag \aaa_k +\sum_{k=1}^3\frac{\omega_{b_k}}{2} \sss_z^k.
$$
The considered errors refer to bit-flips occurring in this rotating frame. In this subsection, we restrict ourselves to the case where such an error only occurs on qubit 1 and at a rate $\gamma_x$. Therefore, we need only a single resonator to perform the correction. 

\begin{figure}[htbp]
\begin{floatrow}
         \subfloat[]{
                \hbox{\hspace{-1.25em}\includegraphics[width=\columnwidth]{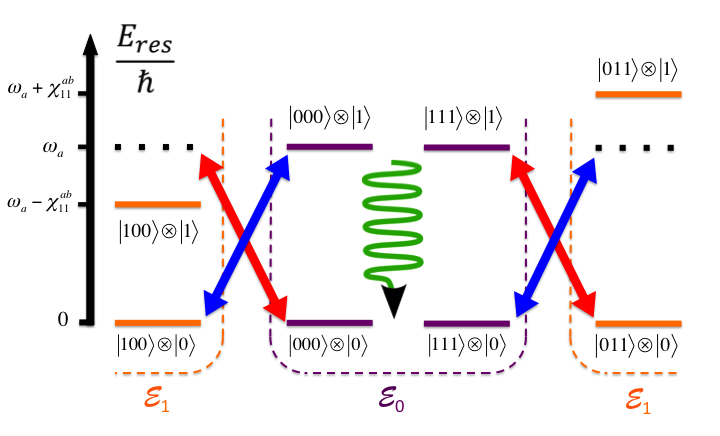}}
                \label{fig:energydiagram}}
 \end{floatrow}
 \begin{floatrow}
         \subfloat[]{
                \hbox{\hspace{-0.2em}\includegraphics[width=1\columnwidth]{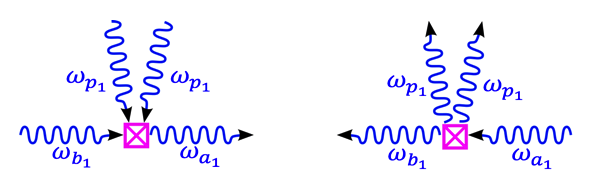}}
                \label{fig:red22}}
 \end{floatrow}  
 \begin{floatrow}
         \subfloat[]{
                \hbox{\hspace{-0.2em}\includegraphics[width=1\columnwidth]{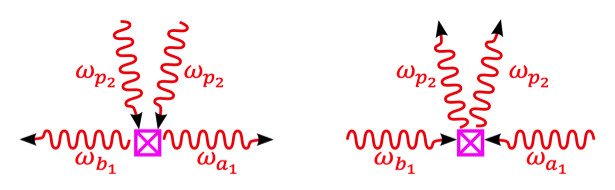}}
                \label{fig:green22}}
 \end{floatrow}  
\caption{(a): Energy-level diagram of the resonator only (the qubits energies are not represented here) as a function of the joint-state of the qubits-cavity system. As explained through Subsection~\protect\ref{sec:level3A}, the resonator never gets populated beyond Fock state $\ket{1}_A$ and therefore, we restrict the diagram to the space spanned by Fock states $\ket{0}_A$ and $\ket{1}_A$. Blue (resp. red) straight-line arrows indicate couplings between two states induced by the pump at frequency $\omega_{p_1}$ (resp. $\omega_{p_2}$). Wavy arrow indicates a common decay channel due to the decay of the single photon in the ancillary resonator. (b): Four-wave mixing process induced by the pump at frequency $\omega_{p_1}$ where two pump photons convert a single qubit excitation to a photon of the resonator (left). This process occurs along with its reverse transformation (right). (c): Four-wave mixing process induced by the pump at frequency $\omega_{p_2}$ where two pump photons create an excitation in both the qubit and the resonator (left). This process occurs along with its reverse transformation (right).}
\end{figure}

Before getting to the details of the scheme, let us provide an intuitive picture (Figure~\ref{fig:energydiagram}). Starting from a superposition state $(c_0\ket{000}+c_1\ket{111})\in\EEE_0$ (while the cavity mode is in the vacuum state $\ket{0}_A$), and after an eventual bit-flip error of the first qubit, the system ends up in the state $(c_0\ket{100}+c_1\ket{011})\in\EEE_1$.  Applying  microwave drives of fixed and well-chosen frequencies, we induce an effective transition between the states $\ket{100}\otimes\ket{0}_A$ and $\ket{000}\otimes \ket{1}_A$ and another one between the states $\ket{011}\otimes\ket{0}_A$ and $\ket{111}\otimes \ket{1}_A$. Note that, through the choice of the drive frequencies, these transitions are turned on in a selective manner, only when the three qubits lie in the manifold $\EEE_1$.  Moreover, by fixing the amplitudes of the drives, these transitions which are illustrated by straight-line arrows in Figure~\ref{fig:energydiagram}, will conserve the initial superposition $(c_0\ket{100}+c_1\ket{011})\otimes \ket{0}_A$ producing the state $(c_0\ket{000}+c_1\ket{111})\otimes \ket{1}_A$. Now a rapid decay of the ancilla resonator resets its state to the vacuum and projects the three-qubit system to the code space. Through the following paragraphs, we will detail the  ingredients of this protocol.

\textit{Three-qubit manifold selectivity} - We apply two continuous-wave (CW) microwave drives at frequencies $\omega_{p_1} = {|\omega_{a_1} -\omega_{b_1}|}/{2} $ and $\omega_{p_2} = {(\omega_{a_1} +\omega_{b_1})}/{2} $, and amplitudes $\epsilon_{p_1}$ and  $\epsilon_{p_2}$. These drives are far detuned from all resonance frequencies and act as stiff pumps in a parametric procedure. As illustrated in Figure~\ref{fig:red22}, two pump photons at frequency $\omega_{p_1}$ convert an excitation in qubit 1 to an excitation of the resonator. In the same way (Figure~\ref{fig:green22}), two pump photons at  frequency $\omega_{p_2}$ create, simultaneously, an excitation both in the qubit and in the resonator. These processes happen in a coherent manner and the oscillation rate and phase can be tuned by adjusting the pumps amplitudes and phases. In particular, we choose these amplitudes and phases to ensure the same rate and  phase for both oscillations, leading to an effective Hamiltonian of the form
\begin{align}
\label{eq:completeH}
 \bold{H}_{\text{eff}}(t) &= -\aaa_1^{\dagger}\aaa_1(\frac{\chi^{ab}_{11}}{2}\sss_z^1+\frac{\chi^{ab}_{12}}{2}\sss_z^2+\frac{\chi^{ab}_{13}}{2} \sss_z^3) \notag
 \\
 & + \frac{\Omega_{p_1}}{2}(\sss_{+}^1\aaa_1+c.c.) + \frac{\Omega_{p_2}}{2}(\sss_{-}^1\aaa_1+c.c.).
 \end{align} 
 Here, the second line of the Hamiltonian is derived from the fourth-order terms of the cosine in~\eqref{eq:bareH} and after applying a rotating wave approximation (RWA). The Rabi frequency $\Omega_{p_j}$ are given by 
 
 \begin{align} \label{eq:omegaexpression}
 \Omega_{p_j} = \sqrt{\chi_{11}^{aa}\chi_{11}^{ab}}\Big{|}{\frac{\epsilon_1^{a,j}}{\omega_{a_1}-\omega_{p_j}}}\Big{|}^2.
\end{align}
 
The amplitudes of the pumps $\epsilon_{1}^{a,1}$ and $\epsilon_{1}^{a,2}$ are chosen such that $\Omega_{p_1}$ and  $\Omega_{p_2}$ are real quantities both equal to $\Omega_p$. Note that, for simplicity sakes, we have neglected the other self-Kerr and cross-Kerr terms $\chi^{aa}_{jk}$ and $\chi^{bb}_{jk}$. We will discuss their effect at the end of the next subsection.
 
 Taking into account the dispersive shifts $\chi^{ab}_{1j}$, the  pump tone $\omega_{p_1}$  only affects the transition between $\ket{011}\otimes\ket{0}_A$ and $\ket{111}\otimes\ket{1}_A$. In the same manner, the pump tone $\omega_{p_2}$ only affects the transition between $\ket{100}\otimes\ket{0}_A$ and $\ket{000}\otimes\ket{1}_A$. In particular, as illustrated in Figure~\ref{fig:energydiagram}, the manifold $\EEE_0\otimes\ket{0}_A$ is left untouched: the transitions  $\ket{000}\otimes\ket{0}_A \leftrightarrow  \ket{100}\otimes\ket{1}_A$ and $\ket{111}\otimes\ket{0}_A \leftrightarrow  \ket{011}\otimes\ket{1}_A$ are detuned by $\pm \chi^{ab}_{11}$ from twice the pump tones. Therefore, the strong dispersive coupling ensured by~\eqref{eq:strongdisp} provides the selectivity of the manifold $\EEE_1$ in the correction procedure. 

One can note that, during the correction procedure, the resonator is only populated when the three-qubit system is in the manifold $\EEE_0$. By the assumption~\eqref{eq:deg}, in such a case the resonator's frequency is given by $\omega_{a_1}$ independently of the states $\ket{000}$ or $\ket{111}$ of the three qubits. This degeneracy ensures that the outgoing photons of the resonator do not reveal any further information about the superposition between these two states.  

Finally, the dissipation of the ancilla resonator projects the three-qubit state to the code space $\EEE_0$ and resets the resonator to its vacuum state. Evacuating the information entropy, this ensures the irreversibility of the transition from $\EEE_1$ to $\EEE_0$.

\textit{Effective model - } Throughout the rest of this subsection, we provide a reduced model and derive an effective correction rate.  We start by moving into the rotating frame of 
$\bold{H_{\text{disp}}}=-\aaa_1^{\dagger}\aaa_1(\frac{\chi^{ab}_{11}}{2}\sss^1_z+\frac{\chi^{ab}_{12}}{2}\sss^2_z+\frac{\chi^{ab}_{13}}{2} \sss^3_z) $. The resonance frequencies being well-resolved~\eqref{eq:strongdisp}, we apply the RWA, removing highly oscillating terms at frequencies of order $\chi^{ab}_{jk}$. Furthermore, choosing $\Omega_p<\kappa_1$, we can adiabatically eliminate the low-Q resonator mode to achieve the following effective master equation:
\begin{align}\label{eq:reduced}
\begin{array} {lcl}
\frac{d\rrr}{dt} = \Gamma_c \mathcal{D} [\bold{c}_1](\rrr) 
+ \frac{\gamma_x}{2} \mathcal{D} [\sss^1_x](\rrr).
\end{array}
\end{align}
In this master equation, $\mathcal{D}[\bold{o}](\rrr)=\bold{o}\rrr\bold{o}^\dag-1/2(\bold{o}^\dag\bold{o}\rrr+\rrr\bold{o}^\dag\bold{o})$. While the second Lindblad term formulates the bit-flip errors of the qubit 1, the first term represents the effective error correction. Here, the induced correction operator,  $\bold{c}_1$, is given by
\begin{equation*}
\bold{c}_1=\ket{000}\bra{100}+\ket{111}\bra{011}=\sss_-^1\boldsymbol{\Pi}^{23}_{\ket{00}}+\sss_+^1\boldsymbol{\Pi}^{23}_{\ket{11}}.
\end{equation*}
where $\boldsymbol{\Pi}^{23}_{\ket{00}}$ (resp. $\boldsymbol{\Pi}^{23}_{\ket{11}}$) is the projection operator of the second and third qubit on the state $\ket{00}$ (resp. $\ket{11}$).
Moreover, $\Gamma_c$ represents the effective correction rate and is well approximated by $\Gamma_c \approx \frac{\Omega_p^2}{\kappa}$.

\begin{figure}
\hbox{\hspace{-0.75em}\includegraphics[width=1.1\columnwidth]{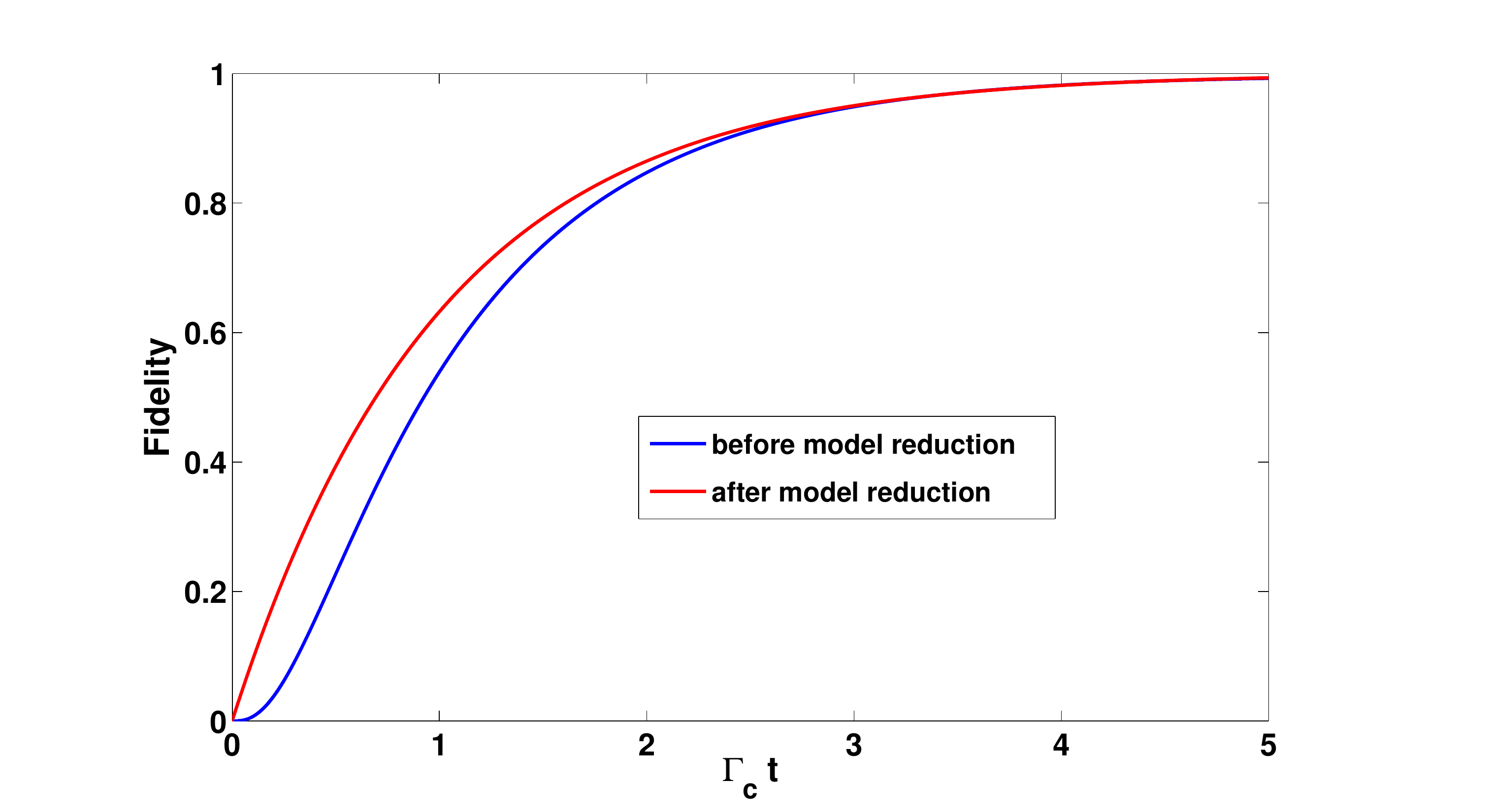}}
\caption[]{Autonomous QEC when the system is initialized in the corrupted state $(\ket{100}-i\ket{011})/\sqrt{2} \in \EEE_1$ and no additional errors occur ($\gamma_x=0$). The red curve illustrates  the fidelity to the state $(\ket{000}-i\ket{111})/\sqrt{2} \in \EEE_0$, for the model given by the Hamiltonian~\eqref{eq:completeH} (before model reduction). The blue curve represents this fidelity for the reduced model given by~\eqref{eq:reduced}. Other parameters have been set to ${\chi_{11}^{ab} = - 20\kappa,\chi_{12}^{ab}=10\kappa,\chi_{13}^{ab} = 10\kappa, \text{and}~\Omega_p = 0.3 \kappa}$, giving rise to $\Gamma_c = 0.09 \kappa$.}
\label{fig:singlequbit}
\end{figure}

The simulation of Figure~\ref{fig:singlequbit} illustrates the performance of this correction protocol. Starting from a corrupted state  $(\ket{100}-i\ket{011})/\sqrt{2}\in\EEE_1$, and neglecting further bit-flip errors ($\gamma_x=0$), we simulate the system's dynamics before and after the model reduction. By plotting the fidelity $F(t)=\bra{\psi_0}\rrr(t)\ket{\psi_0}$ with respect to the state ${\psi_0}=(\ket{000}-i\ket{111})/\sqrt{2}\in\EEE_0$, we observe that the dynamics is well described by the reduced model and that the correction happens at the predicted rate $\Gamma_c$.  

\subsection{\label{sec:level3B}Correction on  three qubits}

Now, we consider the case where each qubit $j$ can independently undergo a bit-flip error at a rate of $\gamma_x^j$. Similarly to the previous subsection, we apply two pumps at frequencies $\omega^j_{p_1} = (\omega^j_{a} +\omega^j_{b})/{2} $ and $\omega^j_{p_2} = |\omega^j_{a}-\omega^j_{b}|/{2}$, both associated to each qubit. Following the derivation of~\eqref{eq:completeH}, this leads to the following effective Hamiltonian 
\begin{align}
\label{eq:completeHthreequbits}
 &\bold{H}_{\text{eff}}(t)=  -\sum\limits_{j=1,2,3} \aaa_j^{\dagger}\aaa_j(\frac{\chi^{ab}_{j1}}{2}\sss^1_z+\frac{\chi^{ab}_{j2}}{2}\sss^2_z+\frac{\chi^{ab}_{j3}}{2}\sss^3_z) 
 \notag\\
 &  +\frac{\Omega_p^j}{2}\sum\limits_{j=1,2,3}(\sss_{+}^j\aaa_j+c.c.)+\frac{\Omega_p^j}{2}\sum\limits_{j=1,2,3}(\sss_{-}^j\aaa_j+c.c.),
 \end{align}
 
 where the $\Omega_p^j$'s are given by expressions similar to~\eqref{eq:omegaexpression}.It is straightforward from Subsection~\protect\ref{sec:level3A} that the reduced dynamics is given by
\begin{equation}
 \label{eq:adiabaticthree}
\frac{d\rrr}{dt} = \sum\limits_{j=1,2,3} \Gamma_c^j \mathcal{D}[\bold{c}_j](\rrr) 
+ \frac{\gamma_x^j}{2} \mathcal{D}[\sss^j_x](\rrr),
\end{equation}
where 
\begin{align*}
\bold{c}_1=\sss_-^1\boldsymbol{\Pi}^{23}_{\ket{00}}+\sss_+^1\boldsymbol{\Pi}^{23}_{\ket{11}}&,~
\bold{c}_2=\sss_-^2\boldsymbol{\Pi}^{13}_{\ket{00}}+\sss_+^2\boldsymbol{\Pi}^{13}_{\ket{11}},\\
\bold{c}_3=\sss_-^3\boldsymbol{\Pi}^{12}_{\ket{00}}&+\sss_+^3\boldsymbol{\Pi}^{12}_{\ket{11}},
\end{align*}
and
$$
\Gamma_{c}^j\approx \frac{|\Omega_{p}^j|^2}{\kappa_j}.
$$
\textit{Effect of other self-Kerr and cross-Kerr terms - } Through the analysis of Subsections~\protect\ref{sec:level3A} and~\protect\ref{sec:level3B}, we have neglected the effect of higher order couplings between various modes as presented in the third and fourth lines of Hamiltonian~\eqref{eq:bareH}. Here, we illustrate that these terms can be fully taken into account without any changes in the performance of the protocol. We only require to slightly modify the pump frequencies. 

These higher order contributions could be considered in two parts. First, the self-Kerr terms and the cross-Kerr terms between the resonator modes 
$$
- \sum_{k} \chi_{kk}^{aa}{\aaa_k}^{\dagger 2}\aaa_k^2- \sum_{j\neq k} \chi_{jk}^{aa }{\aaa_j}^{\dagger}\aaa_j{\aaa_k}^{\dagger}\aaa_k$$
do not affect the dynamics. Indeed, the self-Kerr terms vanish as these modes are never populated beyond a single photon. Similarly, the cross-Kerr terms can be neglected since two resonator modes are never populated simultaneously. 

Next, the cross-Kerr terms between the qubit modes
$$
- \sum_{j\neq k} \chi_{jk}^{bb} \sss^j_z \sss^k_z
$$
yield an identical energy shift to each two states in an error subspace $\EEE_j$. Modifying slightly the pump tones  to take into account these energy shifts, we will get the same effective Hamiltonian (modulo the addition of the above self-Kerr and cross-Kerr terms between resonator modes) as in~\eqref{eq:completeHthreequbits}. More precisely, the modified pump frequencies, associated to qubit 1, are given by\small
\begin{equation}\label{eq:pumptone}
\tilde \omega^1_{p_1}=\frac{\omega_{a_1}+\omega_{b_1}}{2}-\chi^{bb}_{12}-\chi^{bb}_{13},~
\tilde \omega^1_{p_2}=\Big|\frac{\omega_{a_1}-\omega_{b_1}}{2}-\chi^{bb}_{12}-\chi^{bb}_{13}\Big|.
\end{equation}\normalsize
Similar modifications need to be applied to other pump tones.

\subsection{Summary of QEC protocol and numerical simulations}\label{ssec:summary}
Through this subsection, we provide a summary of the requirements for our QEC scheme (presented in previous sections) and we realize numerical simulation to illustrate its performance. We couple three qubits  to three low-Q resonator modes as in Figure~\ref{fig:fancyfig} and we assume the following separation of time-scales:
$$
\gamma_x^j\ll\kappa_{j'}\ll\chi_{k,k'}^{ab},\quad j,k,j',k'=1,2,3. 
$$
We further assume the symmetry assumption
\begin{equation}\label{eq:symm}
\sum_{k=1,2,3}\chi^{ab}_{jk}=0,\quad j=1,2,3.
\end{equation}
As it will be seen through the next section, this can be relaxed to $\sum_{k=1,2,3}\chi^{ab}_{jk}\sim \gamma_x^j$. Such a symmetry should be achievable by fine tuning the frequencies of the qubits. 

Now, we apply six off-resonant CW drives of frequencies $\tilde\omega^j_{p_1,p_2}$ given by~\eqref{eq:pumptone} (with adjusted phases and amplitudes), acting as stiff pumps. This yields an effective master equation of the form
\begin{align} \label{eq:3qubitcorrection}
\frac{d\rrr}{dt}&=-i[\bold{H}_{\text{eff}},\rrr]+\sum\limits_{j=1,2,3}\frac{\gamma_x^j}{2}\DDD[\sss_x^j]\bold{\rho}+\sum\limits_{j=1,2,3}{\kappa_j}\DDD[\aaa_j]\rrr,\notag\\
\bold{H}_{\text{eff}}&=-\sum\limits_{j=1,2,3} {\aaa_j}^{\dagger}\aaa_j(\frac{\chi^{ab}_{j1}}{2}{\sss^1_z}+\frac{\chi^{ab}_{j2}}{2}{\sss^2_z}+\frac{\chi^{ab}_{j3}}{2} {\sss^3_z}) 
 \notag\\
 &  +\frac{\Omega_p^j}{2}\sum\limits_{j=1,2,3}({\sss_{+}^j}{\aaa_j}+c.c.)+\frac{\Omega_p^j}{2}\sum\limits_{j=1,2,3}({\sss_{-}^j}{{\aaa_j}}+c.c.)\notag\\
 &- \sum_{j} \chi_{jj}^{aa}{{\aaa_j}^{\dagger 2}\aaa_j^2}- \sum_{j\neq k} \chi_{jk}^{aa }{{\aaa_j}^{\dagger}\aaa_j}{{\aaa_k}^{\dagger}\aaa_k},
\end{align}
where the $\Omega_p^j$'s are given by 

 \begin{equation}\label{omegap}
 \Omega_p^j = \sqrt{\chi_{11}^{aa}\chi_{11}^{ab}}\Big|{\frac{\epsilon_j^{a,1}}{\omega_{a}^j-\tilde\omega_{p_1}^j}}\Big|^2= \sqrt{\chi_{11}^{aa}\chi_{11}^{ab}}\Big|{\frac{\epsilon_j^{a,2}}{\omega_{a}^j-\tilde\omega_{p_2}^j}}\Big|^2.
 \end{equation}

\begin{figure}
	
		\hbox{\hspace{-1.0em}\includegraphics[width=1.1\columnwidth]{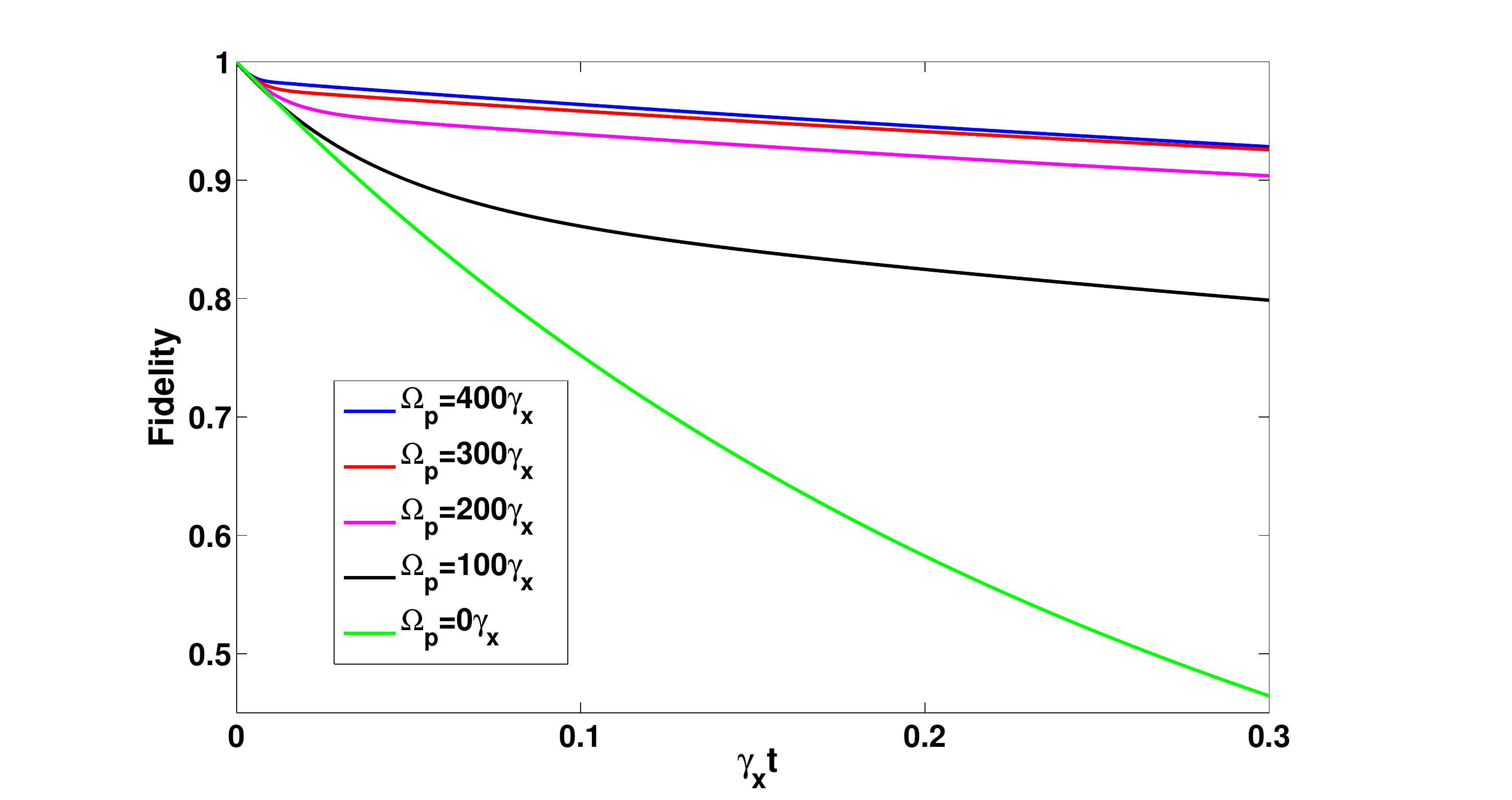}}
	\caption[purity]{Simulation of model~\eqref{eq:3qubitcorrection} for the QEC scheme when the system is initialized in the state ${(\ket{000}-i\ket{111})/\sqrt{2}\in\EEE_0}$. The curves illustrate the fidelity to the initial state for different values of ${\Omega_p}$ (all $\Omega_p^j$ are taken to be identical to $\Omega_p$). Each qubit suffers from bit-flip errors at a rate $\gamma_x^j =\gamma_x$, and the dissipation rates of the resonators are taken to be equal and set to $\kappa = 500\gamma_x$. The dispersive couplings of qubit 2 and qubit 3, $\chi_{j2}^{ab}$ and $\chi_{j3}^{ab},j=1,2,3,$ are chosen to be positive and satisfy ${\chi_{j2}^{ab}=\chi_{j3}^{ab}=-\chi_{j1}^{ab}/2}$. The coupling strengths of qubit 1, $\chi_{j1}^{ab}$, are equal and swept so as to keep the ratio ${\Omega_p/\chi_{j1}^{ab}=10^{-2}}$ constant. The cross-Kerr coefficients between resonators are taken to be $\chi_{jk}^{aa}=\chi_{jk}^{ab}/100$.}
	\label{fig:purity}
\end{figure}

In Figure~\ref{fig:purity}, we simulate the above master equation. We fix the decay rates of the low-Q modes to be $\kappa=500\gamma_x$ and we sweep the dispersive shift strengths $\chi^{ab}_{jk}$ and the pump-induced transition rates $\Omega_p^j$ (keeping their ratio constant). The system is initialized in $\ket{\psi_0}=(\ket{000}-i\ket{111})/\sqrt{2}$, the -1 eigenstate of the logical operator ${\sss_y^L}=-\sss_y^1 \sss_y^2 \sss_y^3$. By tracing the fidelity $F=\bra{\psi_0}\bold{\rho}\ket{\psi_0}$ with respect to this initial state, we show that the autonomous correction enhances significantly the lifetime of the encoded state. In particular, after a time of order $1/(\gamma_x^1+\gamma_x^2+\gamma_x^3)$, we  maintain a fidelity in excess of $90\%$.  Besides, we observe that while increasing $\Omega_p^j$ improves the correction rate as predicted by formula $\Gamma^j_c=|\Omega^j_p|^2/\kappa$, this rate is saturated when $\Omega^j_p$ approaches $\kappa$. This corresponds to the fact that the entropy cannot be  evacuated at a rate faster than $\kappa$. This saturation limit can be enhanced by increasing the decay rate of the low-Q mode while the qubit decay rates remain constant. While in principle this separation of decay rates is usually limited by Purcell effects, in practice we can design Purcell filters to overcome this limitation~\cite{Shankar-Nature-2013,ZakiCats2}. Note that the second order effect of the highly oscillating terms neglected in the RWA of Subsection~\eqref{sec:level3A} induces an extra phase shift between the two logical states. This phase shift is however deterministic and does not corrupt the encoded quantum information. In the above simulations we take this deterministic phase into account for the calculation of the relative fidelity.

\section{\label{sec:level4}Effective decoherence rate after QEC}

Our first order QEC protocol is not capable of correcting two errors occurring within a time given by the inverse of the correction rate. Instead, it will recover a wrong state inside the code space corresponding to a bit-flip of the logical qubit. This leads to an effective second-order decay rate given by (see~\cite{nielsen00}, Chapter 10)
$$
\Gamma_{\text{eff}}^{2\text{nd}}=3\gamma_x^2/\Gamma_c.
$$  
This decoherence rate corresponds to the ideal case where all the model reductions of Subsection~\protect\ref{sec:level3A} are exact. In practice, one needs to take account further decoherence rates induced by the imperfection of the RWA and the eventual breakdown of symmetry~\eqref{eq:symm}. Through this section, we present the requirements to reduce the major such effects to the same order as the above effective decoherence rate $\Gamma_{\text{eff}}^{2\text{nd}}$. 

\textit{Imperfect manifold selectivity - }
A major requirement for the protocol to perform as predicted, is that the pump tones $\tilde\omega_{p_{1,2}}^j$ induce oscillations, only, between the manifolds $\EEE_j\otimes \ket{0}_j$ ($\ket{0}_j$ corresponds to the vacuum state of resonator $j$) and $\EEE_0\otimes \ket{1}_j$. In particular, the manifold $\EEE_0\otimes\ket{0}_j$ should remain untouched. As stated in Subsection~\protect\ref{sec:level3A}, this manifold selectivity is provided by the fact that transitions between $\EEE_0\otimes\ket{0}_j$ and $\EEE_j\otimes\ket{1}_j$ are off-resonant by $\pm\chi^{ab}_{jj}$ (see Figure~\ref{fig:energydiagram}). However, in practice, this undesired manifold $\EEE_j\otimes\ket{1}_j$ gets slightly populated due to the finite ratio $\chi^{ab}_{jj}/\kappa_j$ between the detuning and the linewidth. This resonator $j$ eventually leaks out its photon carrying information about the logical superposition. This leads to an effective dephasing rate given by 
\begin{equation}\label{eq:gamselect}
\Gamma_{\text{eff}}^{\text{select}} = \sum_{j=1,2,3}\kappa_j\frac{|\Omega_p^j|^2}{|\chi_{jj}^{ab}|^2+|\kappa_j|^2}.
\end{equation}
This rate could be understood by the fact that the average population of the undesired manifold $\EEE_j\otimes\ket{1}_j$ due to the detuned pumps is given by ${|\Omega_p^j|^2}/({|\chi_{jj}^{ab}|^2+|\kappa_j|^2})$.
 
\textit{Symmetry breakdown - } As stated in Subsection~\ref{sec:level3A}, in order to not leak out any  information on a given superposition between the states $\ket{000}$ and $\ket{111}$, we need to ensure a symmetry assumption given by relation~\eqref{eq:symm}. Here, we assume that such an assumption is not perfectly satisfied and we quantify its major contribution to an induced decoherence rate. 

This major effect is due to the fact that whenever the system undergoes a bit-flip (rate $ \gamma_x^j$), the protocol performs a transition from $\EEE_j\otimes\ket{0}_j$ to $\EEE_0\otimes\ket{1}_j$. The three-qubit system then accumulates a relative phase (rate $\Big|\sum\limits_{k=1,2,3}\chi^{ab}_{jk}\Big|$) before the photon is lost (time of order $1/\Gamma_c^j$). This induces an effective dephasing rate of order 
\begin{equation}\label{eq:gamsym}
\Gamma_{\text{eff}}^{\text{sym}}\sim \sum\limits_{j=1,2,3}\gamma_x^j \frac{\big|\sum\limits_{k=1,2,3}\chi^{ab}_{jk}\big|}{\Gamma_c^j}. 
\end{equation}
To sum up, we provide the requirements to reduce the effect of these imperfection-induced decoherence rates to the same order as the second order bit-flip errors. Increasing the pump powers ($\Omega_p$'s of the same order as $\kappa$'s), we saturate the correction rate $\Gamma_c$ to a rate of order $\kappa$. Then the rate $\Gamma_{\text{eff}}^{\text{select}}$ becomes of the same order as $\Gamma_{\text{eff}}^{2\text{nd}}$, whenever 
$$
\frac{\kappa_j}{\gamma^j_x}\lesssim \frac{\chi^{ab}_{jj}}{\kappa_j}.
$$
Similarly, for the rate $\Gamma_{\text{eff}}^{\text{sym}}$, we need to take 
$$
\mid\sum\limits_{k=1,2,3}\chi^{ab}_{jk}\mid\lesssim\gamma_x^j.
$$

\section{\label{sec:level5}Towards a simplified implementation}

\begin{figure}

\begin{floatrow}
         \subfloat[]{
                \hbox{\includegraphics[width=0.5\columnwidth]{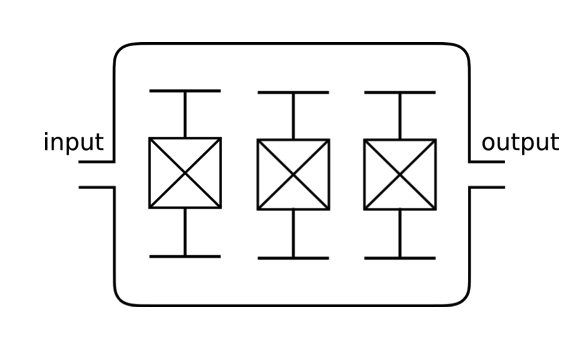}}
                \label{fig:setupone}}       
                   \hspace{-1cm}
	\subfloat[]{
                \includegraphics[width=0.5\columnwidth]{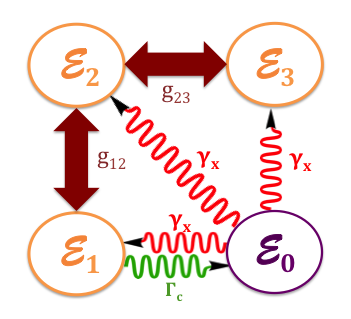}
                \label{fig:schemediagram}}
                
 \end{floatrow}
                
                \caption{(a) Illustration of the simplified scheme, where three superconducting qubits are coupled to a single resonator. (b) A diagram of the single-resonator scheme. Bit-flip errors occurring at a rate $\gamma_x$ induce jumps from the coding subspace $\EEE_0$ to the error subspaces $\EEE_j$. Applying two stiff pumps on the resonator, any bit-flip error is eventually mapped onto a bit-flip of first qubit, on which correction takes place.}
 \end{figure}

Through this section, we propose a simplified version of the above protocol that only requires the coupling of the three qubits to a single low-Q resonator (Figure~\ref{fig:setupone}). As explained through Section~\protect\ref{sec:level3A}, using a single resonator and two CW drives at frequencies $\omega_{p_1}=(\omega_a+\omega_{b_1})/2$ and $\omega_{p_2}=|\omega_a-\omega_{b_1}|/2$, one can autonomously correct bit-flip errors occurring on qubit 1. Here, instead of adding extra resonators (acting as correction channels) for the other qubits, we propose to design an effective Hamiltonian which transfers the errors of the other qubits on this first qubit. 

More precisely, we apply two extra CW drives of fixed amplitudes and phases at frequencies $\omega_{p_{12}} = |\omega_{b_1}-\omega_{b_2}|/2 $  and $\omega_{p_{23}} = |\omega_{b_2}-\omega_{b_3}|/2 $ (four stiff pumps in total). These drives acting as stiff pumps induce effective couplings of the form 
$g_{12}(\sss_+^{1} \sss_-^{2} + c.c.)+g_{23}(\sss_+^{2} \sss_-^{3} + c.c.)$ to be added to the Hamiltonian~\eqref{eq:completeH}. As illustrated in Figure~\ref{fig:schemediagram}, the first term maps coherently ${\mathcal{E}}_2$ to ${\mathcal{E}}_1$ and the second term maps ${\mathcal{E}}_3$ to ${\mathcal{E}}_2$. This induces coherent oscillations between the error subspaces $\EEE_1$, $\EEE_2$ and $\EEE_3$. Now taking into account the irreversible correction procedure occurring when the system passes by the manifold $\EEE_1$ (decay from $\EEE_1$ to $\EEE_0$ in Figure~\ref{fig:schemediagram}), we end up correcting all possible errors. 

\begin{figure}
	\centering
		\hbox{\hspace{-0.75em}\includegraphics[width=1.1\columnwidth]{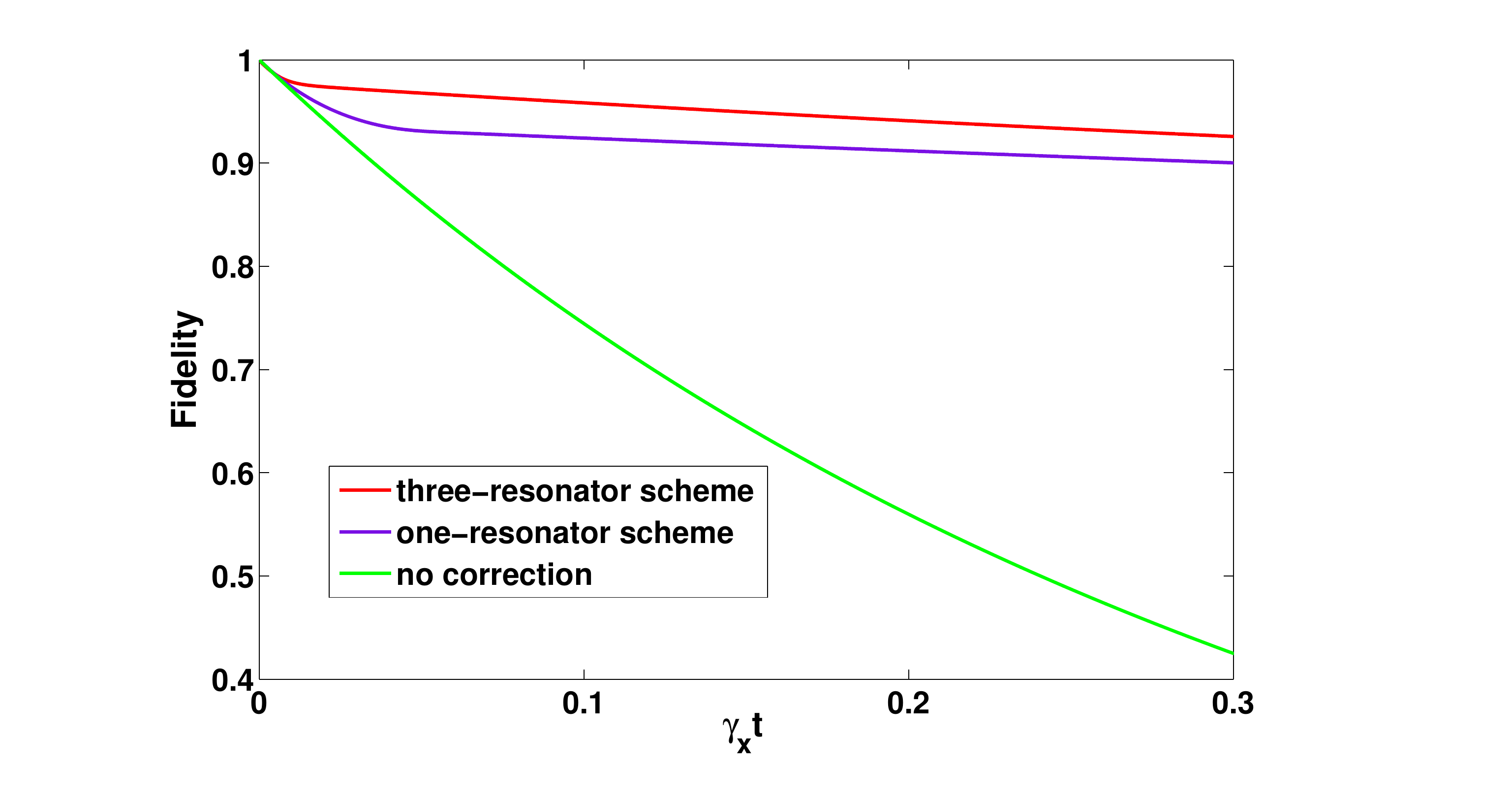}}
	\caption[purity]{QEC protocol using one resonator described by~\eqref{eq:oneresME} when the system is initialized in the state ${(\ket{000}-i\ket{111})/\sqrt{2}\in\EEE_0}$. The purple curve shows the fidelity to the initial state achieved with the single-resonator scheme. In comparison, we also plot the fidelity obtained through the three-resonator scheme presented in Section~\protect\ref{sec:level3} (red curve), and the fidelity without any correction (green curve). Each qubit suffers from bit-flip errors at a rate $\gamma_x$. The decay rate of all resonators is set to $\kappa = 500\gamma_x$ and the transition rates are all given by $\Omega_p = 300\gamma_x$, $g_{12}=\Gamma_c/2$ and $g_{23}=-g_{12}/\sqrt{2}$. The coupling strengths satisfy ${\chi_{j2}^{ab}=\chi_{j3}^{ab}=-\chi_{j1}^{ab}/2}$, $j=1,2,3$, and ${\Omega_p/\chi_{j1}^{ab}=10^{-2}}$.}
	\label{fig:onerespurity}
\end{figure}

The oscillation rates between the error subspaces, $g_{12}$ and $g_{23}$, as well as the associated phases can be tuned by the choice of pump amplitudes and phases. The choice of $g_{12}=\Gamma_c/2$ and $g_{23}=-g_{12}/\sqrt{2}$ corresponds to an optimal effective correction rate. In Figure~\ref{fig:onerespurity}, we simulate such an error correction scheme based on the use of a single resonator and compare it with the previous case of the correction with three resonators. The master equation simulated is given by
\begin{align} \label{eq:oneresME}
\frac{d\rrr}{dt}&=-i[\bold{H}_{\text{eff}},\rrr]+\sum\limits_{j=1,2,3}\frac{\gamma_x^j}{2}\DDD[{\sss_x^j}]\rrr+\kappa\DDD[\bold{a}]\rrr\notag\\
\bold{H}_{\text{eff}}&=-{\aaa}^{\dagger}\aaa(\frac{\chi^{ab}_{11}}{2}{\sss^1_z}+\frac{\chi^{ab}_{12}}{2}{\sss^2_z}+\frac{\chi^{ab}_{13}}{2}{\sss^3_z})  \notag\\
 &  +\frac{\Omega_p}{2}({\sss_{+}^1}\bold{{a}}+c.c.)+\frac{\Omega_p}{2}({\sss_{-}^1}\bold{{a}}+c.c.)\notag\\
 & +g_{12}(\sss_+^{1} \sss_-^{2} + c.c.)+g_{23}(\sss_+^{2} \sss_-^{3} + c.c.).
\end{align}
As can be observed in Figure~\ref{fig:onerespurity}, the effective correction rate is lower for the case of the simplified protocol. This could be understood by the fact that, at each time, we are only able to correct a single error channel. This is to be compared to the  three-resonator protocol, where the three independent error channels are corrected simultaneously. Indeed, the optimal correction rate for this simplified protocol appears to be precisely three times lower than the rate for the three-resonator one. However, for the same reason, the dephasing rate induced by the imperfection in the manifold selectivity (finite ratios $\chi^{ab}_{jk}/\kappa_j$) appears to be at least three times higher for the protocol based on three resonators. This explains the steeper slope of the curve for the three-resonator protocol on the longer time scales.

\section{\label{sec:conclusion}Conclusion}

In conclusion, we have presented a quantum error correction scheme adapted to superconducting circuits that does not require any external feedback loop, but works in an autonomous way through quantum reservoir engineering. The scheme protects a logical qubit encoded in the three-qubit code against bit-flip errors, using three transmon qubits dispersively coupled to three low-Q resonators. {We exploit the strong nonlinearity of the Josephson elements to directly build the feedback loop into the Hamiltonian thus avoiding any need in a directional (non-reciprocal) transmission of quantum information}. We have shown that by applying continuous-wave microwave drives of appropriate and fixed frequencies and amplitudes to this system, the lifetime of an encoded quantum state can be significantly enhanced. More precisely, numerical simulations realized with currently achievable parameters predict a fidelity to the initial state higher than $90\%$ after a time of the same order as the lifetime of the unprotected system. Besides, we have analytically determined that for the scheme to be efficient, we need only certain ratios to be large {in addition to a basic symmetry requirement.} Finally, the hardware equipment needed for the correction scheme can be lightened through the use an alternative scheme, which requires to couple three transmon qubits to only one low-Q resonator at the cost of a slightly slower correction efficiency.

\section*{Acknowledgement}

The authors thank Zaki Leghtas for fruitful discussions on the subject. This work was partially supported by the French ``Agence Nationale de la Recherche" under the project EPOQ2 number ANR-09-JCJC-0070.

\end{document}